\documentclass[11pt]{article}
\usepackage{amsmath,amsfonts, amssymb,graphicx,geometry,authblk,color} 
\UseRawInputEncoding

\graphicspath{{Figures/}}

\title{Multiscale modeling of materials: Computing, data science, uncertainty and goal-oriented optimization}
\author{Nikola Kovachki\footnote{These authors have contributed equally.} , Burigede Liu$^*$, Xingsheng Sun$^*$, Hao Zhou$^*$, \\
Kaushik Bhattacharya\footnote{Corresponding author: bhatta@caltech.edu} , Michael Ortiz, Andrew Stuart}
\affil{California Institute of Technology}

\begin{document}
\maketitle
\begin{abstract}
The recent decades have seen various attempts at accelerating the process of developing  materials targeted towards specific applications.  The performance required for a particular application leads to the choice of a particular material system whose properties are optimized by manipulating its underlying microstructure through processing.  The specific configuration of the structure is then designed by characterizing the material in detail, and using this characterization along with physical principles in system level simulations and optimization.   These have been advanced by multiscale modeling of materials, high-throughput experimentations, materials data-bases, topology optimization and other ideas.  Still,  developing materials for extreme applications involving large deformation, high strain rates and high temperatures remains a challenge.  This article reviews a number of recent methods that advance the goal of designing materials targeted by specific applications.
\end{abstract}

%%%%%%%%%%%%%%%%%%%%%%%%%%%%%%%%%%%%%%%%%%%%%
\section{Introduction}

\begin{figure}
    \centering
    \includegraphics[width=0.9\columnwidth]{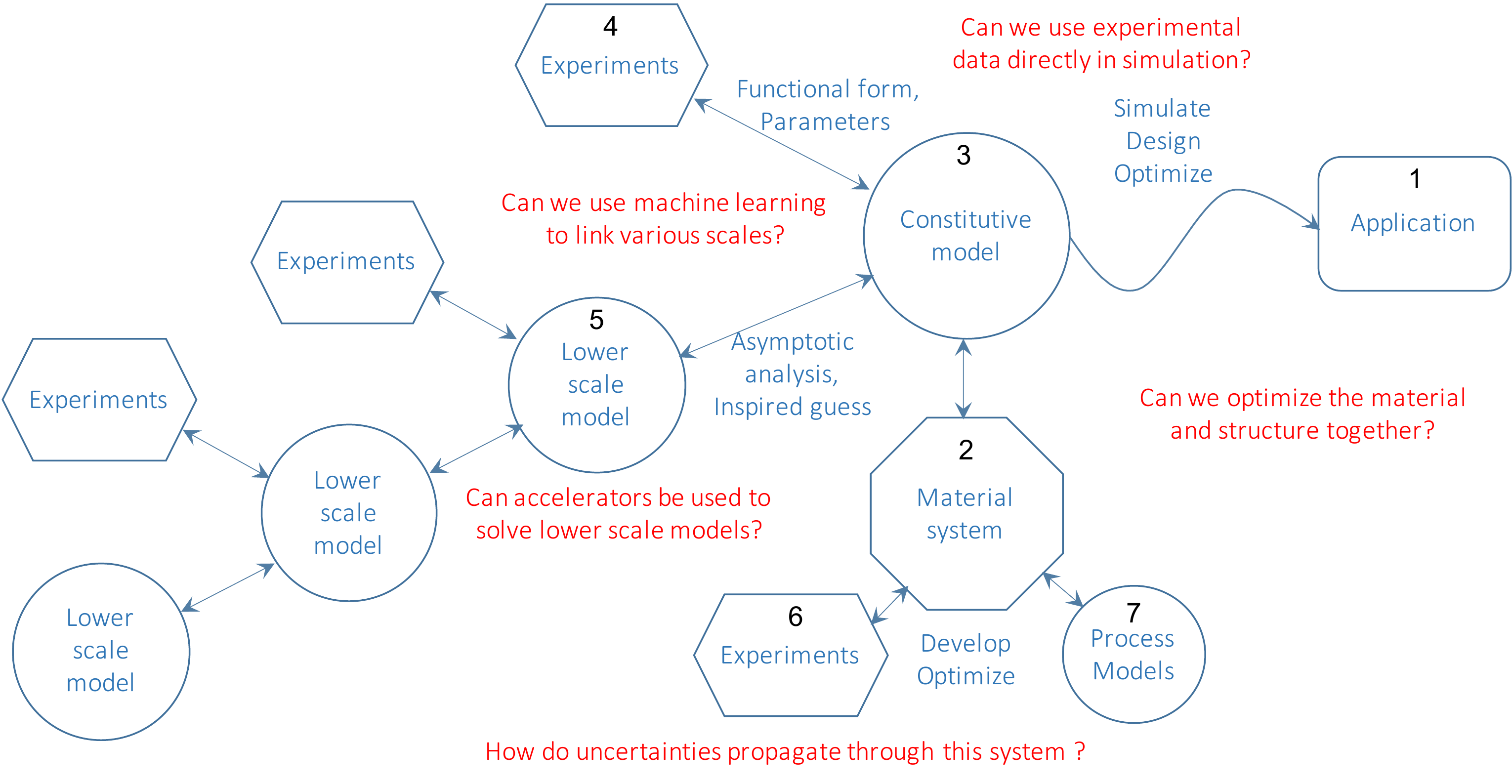}
    \caption{The framework of application-driven materials by design.}
    \label{fig:overview}
\end{figure}

The development of new materials guided by the process-structure-properties-performance paradigm has been the core endeavor of materials science, while the design and optimization of machines by using physical principles and materials characterization have been the core endeavor of mechanical and structural engineering.  These have gradually merged as we seek to develop materials focused on particular applications, especially those involving extreme conditions.  A general framework has emerged and this is described in Figure \ref{fig:overview}.

We seek to design an application (``1'' in Figure \ref{fig:overview}) with a particular material system (``2" in Figure \ref{fig:overview}).  In the established approach of mechanical design, we characterize the material system with a constitutive relation (``3'' in Figure \ref{fig:overview})
and use it along with physical principles (balance laws) to design and optimize an application.    The constitutive relation is intuited from experiments to characterize the material system (``4'' in Figure \ref{fig:overview}), qualitative knowledge from lower scale physics (``5'' in Figure \ref{fig:overview}), symmetry, ease of implementation, history and other considerations.   Multiscale modeling of materials seeks to make  the connection to lower scale models stronger.  The lower scale models are built on the basis of additional experiments at that scale and even lower scale models.  And so on.  Thus, the modeling of materials results in a complex cascade of models at different scales addressing different phenomena at different levels of fidelity.   The material system is itself developed and optimized through a series of experiments (``6'' in Figure \ref{fig:overview}) and process models (``7'' in Figure \ref{fig:overview}).  The process models may also have their own multiscale cascade.  

While this conceptual framework is widely used, the detailed implementation requires expert judgement and decisions due to limitations in the amount of data that is available, limitations in the theoretical and modeling tools and the prohibitive computational cost of a brute force implementation.   \emph{The goal of this paper is to review recent work that provides ideas and tools to implement and exploit this framework for rapidly developing and optimizing materials for particular applications}.

\begin{figure}
    \centering
    \includegraphics[width=0.5\columnwidth]{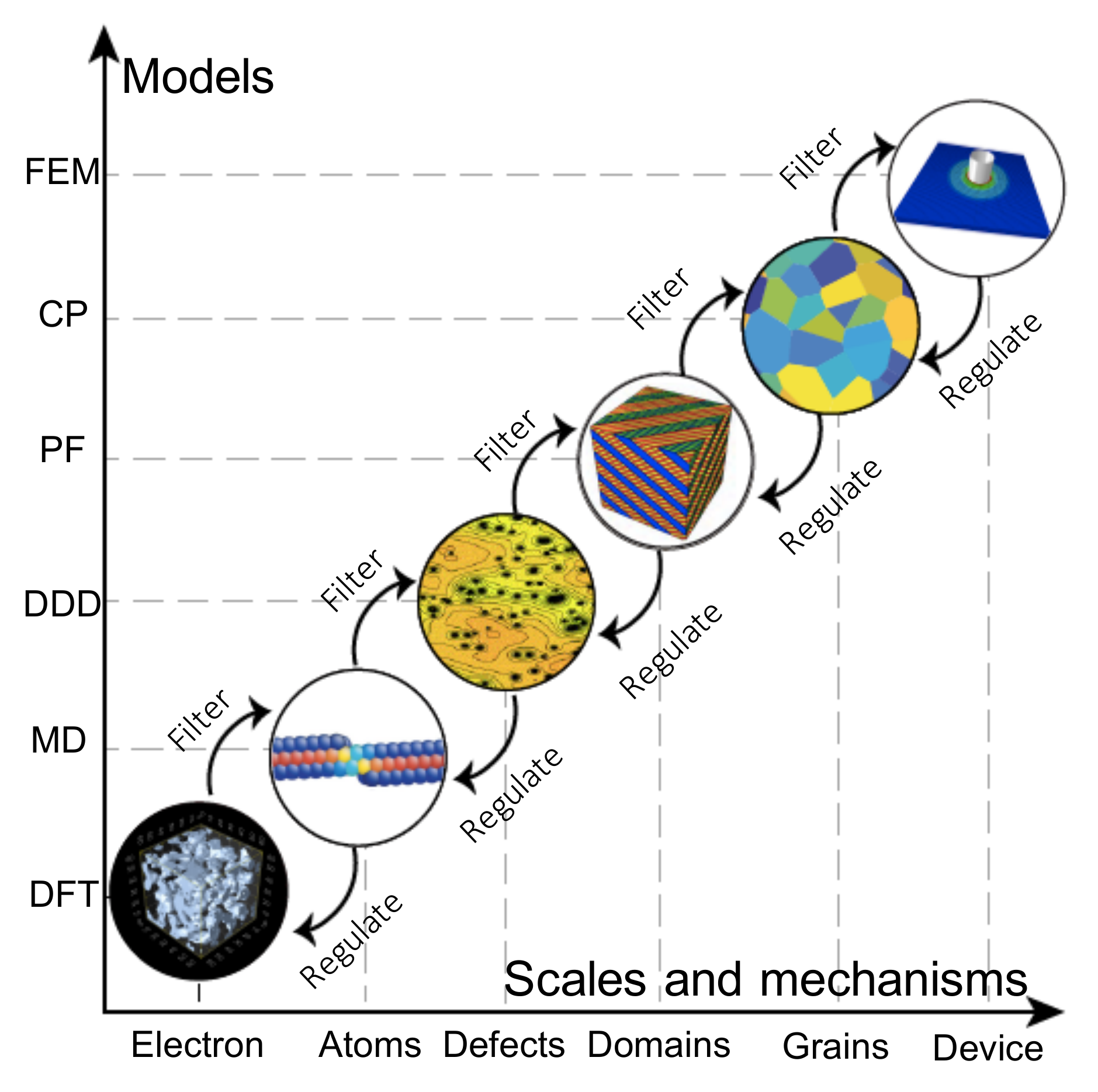}
    \caption{Multiscale modeling of materials is a `divide and conquer' approach to describe the complexity of material behavior.}
    \label{fig:multiscale}
\end{figure}

Underlying the framework described above is the observation that the macroscopic behavior of materials is the end result of  a number of mechanisms that operate across a broad range of disparate scales \cite{p_book_01}.  The paradigm of multiscale modeling seeks to address this complexity using a `divide and conquer' approach shown in Figure \ref{fig:multiscale} \cite{Ortiz:2001,f_book_09,dr_book_11,van2020roadmap}.  The complex range of material behavior is first divided into an ordered hierarchy of scales, the relevant mechanisms at each scale are identified and analyzed using theories/tools based on an individual scale, and the hierarchy is put back together by passing information between scales.    Importantly, the passage of information between scales is pair-wise, with the larger-scale model both modulating (through average kinematic constraints like the boundary conditions) and averaging (the dynamic response like the stress) the smaller-scale model.   The mathematical theory of homogenization \cite{bensoussan2011asymptotic,pavliotis2008multiscale} provides a concrete basis in specialized situations, but the underlying conceptual framework is widely used.

The computational implementation of this framework requires the repeated solution of the models at individual scales.  Recent computing platforms complement the central processing units (CPUs) with massively parallel accelerators like graphics processing units (GPUs)\cite{kirkhwu,exascale}.    Such accelerators contain thousands of processors, but these are not independent.  Instead, they are grouped together in `warps' that share a memory and execute the same instructions but on possible different data (SIMD).  Consequently, they can provide enormous computational power if the calculations are carefully arranged to meet the limitations of the architecture. This raises the first question: \emph{Can we use accelerators to efficiently and rapidly solve models at the individual scales?}  Section \ref{sec:acc} reviews recent work of Zhou and Bhattacharya \cite{zb_acc_21} that describes how such accelerators can be effectively used in the solution of problems of micromechanics.
A key idea in this work is to note that nonlinear partial differential equations that describe micromechanical phenomena come about through a {\it composition} of universal laws of physics (kinematic compatibility and balance of mass, momenta, energy etc.) and the particular constitutive models of the material under study.   The former are nonlocal, but may be interpreted as projections in appropriate function space. The material behavior is nonlinear and may involve time derivatives, but is spatially {\it local}.  Thus each step is amenable to efficient implementation in accelerators. 

We then turn to the issue of bridging two scales.  In the multiscale paradigm, the behavior of each point at each instant of time in the larger scale model has to be informed by a solution of the smaller scale model.  This idea is implemented directly in the concurrent multiscale approaches like FE$^2$ \cite{feyel2000fe2}.  Unfortunately this is prohibitively expensive especially if one has to study multiple scales.   So, a widely used approach is sequential multiscale or parameter passing method where the form of the larger scale model is postulated and the parameters are evaluated from smaller scale simulations (see \cite{cheng2019first,fu2005multiscale,balasubramanian2002elasto} for examples). 
However, this introduces an additional layer of modeling.  All of this leads to the question: \emph{Can we use use techniques from machine learning, such as Gaussian processes, deep neural networks or other supervised learning techniques to link scales?}  Section 3 reviews the recent work of Liu {\it et al.} \cite{letal_arxiv_21} that studies the two scale problem of the response of polycrystalline magnesium to impact loading.  The key idea is to view the lower-scale model (a polycrystalline ensemble described by crystal plasticity augmented to include twinning in this case) as a map from a strain history to a stress history, and to approximate this map using a combination of model reduction and machine learning \cite{bhks_arxiv_20}.

The complexity of the material response leads to uncertainty, and this uncertainty is often the main source of uncertainty in engineering applications.  Therefore, it is important to quantify the integral uncertainties across the various scales in order to identify adequate design margins for the design.  However, the direct estimation of integral material uncertainties is prohibitively expensive.  This leads to the question: \emph{How do uncertainties propagate across scales, and how do we quantify the integral material uncertainty?}  Section \ref{sec:uq} reviews the recent work of Liu {\it et al.} \cite{liu2021hierarchical} that provides the bounds on the integral uncertainty of penetration of a rigid impactor through a polycrystalline magnesium plate.  The key idea is to exploit the hierarchy of multiscale modeling by viewing the model at each scale as a function and the integral response as a composition of the functions.  We can then bound the integral uncertainties using the uncertainty of each individual scale.

A closely related issue is to design materials focussed on particular applications.  This in turn requires an understanding of the sensitivity of the integral response to individual mechanisms -- for example the sensitivity of the ballistic response to the critical resolved shear stress of a particular slip system.  This is not always easy to evaluate, since the mechanism is removed by various scales from the application.  The problem of finding sensitivities is closely related to that of quantifying uncertainties, and we can again take advantage of the hierarchical structure of the multiscale model.

The emergence of full-field diagnostic methods like digital image correlation \cite{mccormick2010digital} and high energy x-ray diffraction microscopy \cite{schwartz2009electron} is leading to a new environment that is rich in experimental data.  This motivates a new question: {\it Can we use experimental data directly in computations without the use of any constitutive models?}   Section \ref{sec:data} summarizes the emergence of a new approach of model-free data-driven simulations, and describes the work of Karapiperis {\it et al.} \cite{karapiperis2021a} studying shear band formation under triaxial compression of sand.  The key idea is to find those stress and strain fields that satisfy the laws of physics (kinematic compatibility and equilibrium) and whose local relationship best approximates the available data.

Finally, the framework described in Figure \ref{fig:overview} first optimizes the material against some overall measure and then optimizes the configuration in which the material is deployed for a given application.  Thermo-mechanical processing of an alloy can control microstructure (grain size, texture and precipitation) which in turn controls the overall properties (toughness and strength).   Alloying can also do the same either directly (solution hardening) or indirectly (by controlling microstructure).  It has long been recognized that doing so can trade one property to another, and this is done against some composite material metric or Ashby plots depending on the material.  The material is then used for the application and the configuration is then optimized.  This raises the question: \emph{Can we optimize the material and the structure simultaneously?}  Section \ref{sec:app} reviews the work of Sun {\it et al.} \cite{sun2021goal} that shows that the simultaneous optimization of a bi-material plate can lead to significantly better ballistic performance compared to a sequential optimization.  The key idea is that not all parts of a structure perform the same function, and different parts of the structure may have different property requirements. 

%%%%%%%%%%%%%%%%%%%%%%%%%%%%%%%%%%%%%%%%%%%%%
\section{Accelerated computations} \label{sec:acc}

Multiscale modeling of materials often requires the repeated solution of models at individual scales.   We focus on  phenomena at the microstructural scale (phase transformations, plasticity, fracture etc.) that are described by continuum models and partial differential equations.  These are typically solved under periodic boundary conditions, and therefore effectively use 
fast Fourier transform (FFT) following Moulinec and Suquet\cite{moulinec1994fast}.  These FFT-based simulations have been effectively applied to a variety of applications (e.g.thermoelasticity~\cite{anglin2014validation}, elasto-viscoplasticity~\cite{lebensohn2016numerical}, dislocations~\cite{berbenni2020fast}, piezoelectric materials~\cite{vidyasagar2017predicting}, shape-memory polycrystals~\cite{bhattacharya2005model}, and crack prediction of brittle materials~\cite{schneider2020fft}).   Various methods to accelerate the convergence of FFT-based methods using Neumann series approximation \cite{monchiet2012polarization,milton2020,moulinec2014comparison,milton2020}
and Fourier-Galerkin  method \cite{vondvrejc2014fft,mishra2016comparative} have been developed.  

While much of the focus has been on CPUs, recent work has turned to GPUs in parts of the algorithm \cite{bertin2018fft,mihaila2014three,knezevic2014high,eghtesad2018spectral}.
Accelerators like GPUs contain thousands of processors, but these are not independent.  Instead, they are grouped together in `warps' that share a memory and execute the same instructions but on possible different data (SIMD).  Consequently, they can provide enormous computational power if the calculations are carefully arranged to meet the limitations of the architecture.
This section describes how accelerators like graphical processing units (GPUs) can be effectively used in rapidly solving such problems drawing from Zhou and Bhattacharya \cite{zb_acc_21}. 

In the models of interest, we describe the state of the solid by a deformation gradient $F$ and a set of internal variables $\lambda$ (phase fraction, plasticity, director field, fracture field etc.).  In the absence of inertia, these are governed by a set of coupled equations:
$$
\nabla \cdot \left( W_F (\nabla u, \lambda, x) \right) = 0, \quad
W_\lambda (\nabla u, \lambda,x) +  D_v(\lambda_t,x) = 0
$$
where $u: \Omega \to {\mathbb R}^3$ is the deformation, $F: \Omega \to {\mathbb R}^{3 \times 3}$ is the deformation gradient, $\lambda: \Omega \to {\mathbb R}^d$ is an internal variable or order parameter, $W: {\mathbb R}^{3\times 3} \times {\mathbb R}^d \times \Omega \to {\mathbb R}$ is the stored (elastic) energy density,  $D: {\mathbb R}^d  \times \Omega \to {\mathbb R}$ is the dissipation potential that governs the evolution of the internal variables and subscripts denote partial differentiation. We adopt an implicit time discretization. We introduce the compatibility condition on $F$ as a constraint and treat it using the augmented Lagrangian method \cite{g_book_15} to obtain the saddle point problem
\begin{equation*}
\min_{u^{n + 1}, \lambda^{n+1}} \max_{\Lambda^{n+1}} \int_\Omega \left( W( F, \lambda, x) + \Delta t  D\left( {\lambda - \lambda^n \over \Delta t},x \right) + \Lambda \cdot (\nabla u - F) + {\beta \over 2} |\nabla u - F|^2   \right) dx
\end{equation*}
for given $\beta > 0$. This problem could be solved via alternating direction method of multipliers({\it ADMM}) \cite{g_book_15} which is an iterative method.  

Given $F^n, \lambda^n, u^n, \Lambda^n$,
\vspace{-0.1in}
\begin{itemize} 
\setlength{\itemsep}{0pt} \setlength{\parskip}{0pt}

\item {\it Step 1: Local problem.} Update $F, n$ by solving at each $x$
\begin{align}
W_F (F^{n+1},\lambda^{n+1},x) - \Lambda^n + \beta (\nabla u^n - F^{n+1}) = 0, \label{eq:eq}\\
W_\lambda (F^{n+1},\lambda^{n+1},x) + \Delta t {\partial D \over \partial \lambda} \left( {\lambda^{n+1} - \lambda^n \over \Delta t} ,x \right) = 0 \label{eq:kin}.
\end{align}
 
\item {\it Step 2: Helmholtz projection.}  Update $u$ by solving $- \Delta u^{n+1} =   \nabla \cdot \left(-  F^{n+1}  + {1 \over \beta} \Lambda^n \right)$.

\item {\it Step 3: Update Lagrange multiplier.}  Update $\Lambda$ as $\Lambda^{n+1} = \Lambda^n + \beta (\nabla u^{n+1} - F^{n+1})$.

\item {\it Step 4: Check for convergence.} Check both primal and dual feasibility:
\begin{equation} \label{eq:check}
    r_p := || \nabla u^{i} - F^{i} ||_{L^2} \le r_p^\text{tolerance}, \quad r_d := \beta || \nabla u^{i+1} - \nabla u^{i} ||_{L^2} \le r_d^\text{tolerance}
\end{equation}
for given $r_p^\text{tolerance}, r_d^\text{tolerance}$. 

\end{itemize}

This method is known to converge under suitable hypothesis on $W,D$ for all $\beta$ sufficiently large \cite{boyd2011distributed}. Step 1 is a local problem, and can be solved trivially in parallel. Note that it is necessary to solve for $F^{n+1}, \lambda^{n+1}$ accurately in some $L^p$, thus we can have poor convergence of a small number of points.  Therefore, the iterations of all points are not impeded by the poor convergence of a few isolated points.  Step 2 leads to a {\it universal} Poisson's equation for which there are a number of effective parallel solvers.  Step 3 is a trivial local update, and step 4 a simple check.  Thus, this iterative algorithm can be implemented in effectively using accelerators like GPUs as we presently demonstrate. 
This iterative method also has a close connection to the physics.  Step 1 is the constitutive update with the Lagrange multiplier converging to the stress, Step 2 is the compatibility equation with the primal convergence in $r_p$ while the dual convergence in  $r_d$ is equivalent to the equilibrium condition.

\begin{figure*}
   	\begin{center}
   		\resizebox{.85\textwidth}{!}{%
   			\includegraphics[height=3cm]{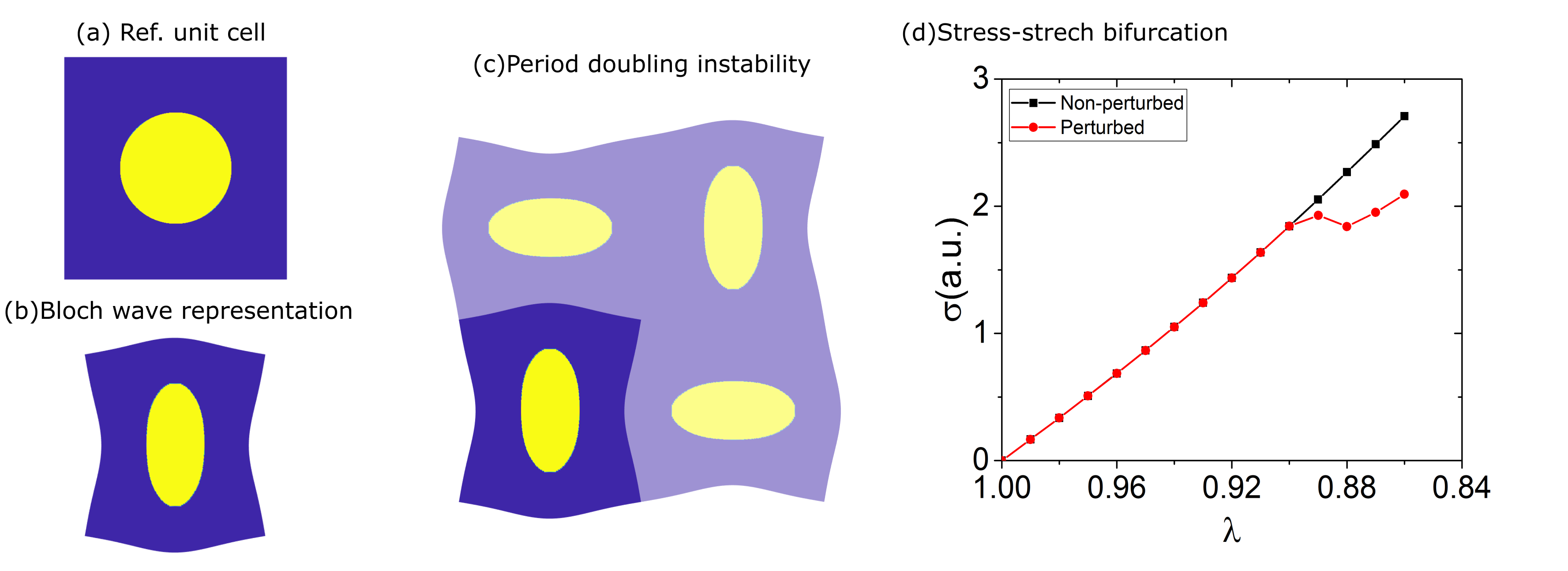}
   		}
   	\end{center}
   	\vspace{-0.3in}
   	\caption{Bifurcation in nonlinear elastic composite.  (a) Unit cell with soft inclusion (yellow) in a hard matrix (blue). (b) Bloch wave instabiity at $\lambda = 0.89$. (c) The period doubling instability on $2 \times 2$ unit cells. (d) Stress-stretch curves with and without instability.}
   	\label{Neo_config}
\end{figure*}

We implement the method on two separate problems.  We start with a problem of finite deformation that involves a bifurcation and has been previously studies using finite element method \cite{triantafyllidis2006failure}. Consider a 2D periodic inclusion of compliant circular particles in a stiff matrix with the unit cell shown in Figure \ref{Neo_config}(a). Both materials are modeled as compressible Mooney-Rivlin materials. As this composite is compressed equi-biaxially, it develops a period doubling instability shown in Figure \ref{Neo_config}(c). The stress-stretch relation shows the bifurcation in Figure \ref{Neo_config}(d). All results are consistent with  previous FE simulations. We also use this example to discuss convergence and scaling.

%We note that the linear stability can also be assessed by the same method.  Recall that stability requires that 
%$\beta = \min_{x \in \mathit{H}^1} \int_{\Omega_0} \overline{v}_{i,j} \mathit{L}_{ijkl} v_{k,l} d X / \int_{\Omega_0} \overline{v}_{i,j} v_{i,j} d X$
%is strictly positive, where $\mathit{L}_{ijkl} = \frac{\partial^2 W}{\partial F_{ij} F_{kl}}$ is the incremental modulus and $v$ ranges from all the admissible incremental displacement field. Further when $\beta \leq 0$, the corresponding minimizer $v$ describes the instability mode. We can rewrite the variational principle describing $\beta$ using augmented Lagrangian and use the algorithm described earlier. We also note that this calculation is possible with the original unit cell by considering the Bloch expression. It simply replaces the gradient $\nabla$ with $(i \omega + \nabla)$ and the Laplacian $\Delta$ in Step 2 with $(i \omega + \nabla)^*  (i \omega + \nabla)$.  

The second concerns Liquid crystal elastomers (LCEs).  These are rubber-like solids where nematic mesogens are incorporated into the polymer chains. The LCE has a spontaneous or stress-free deformation $F$ depending on the mesogens orientation $n$, which is a reorientable unit vector. Therefore, the problem has coupled fields with non-convex energy.

\begin{figure*}
   	\begin{center}
   		\resizebox{.9\textwidth}{!}{%
   			\includegraphics[height=3cm]{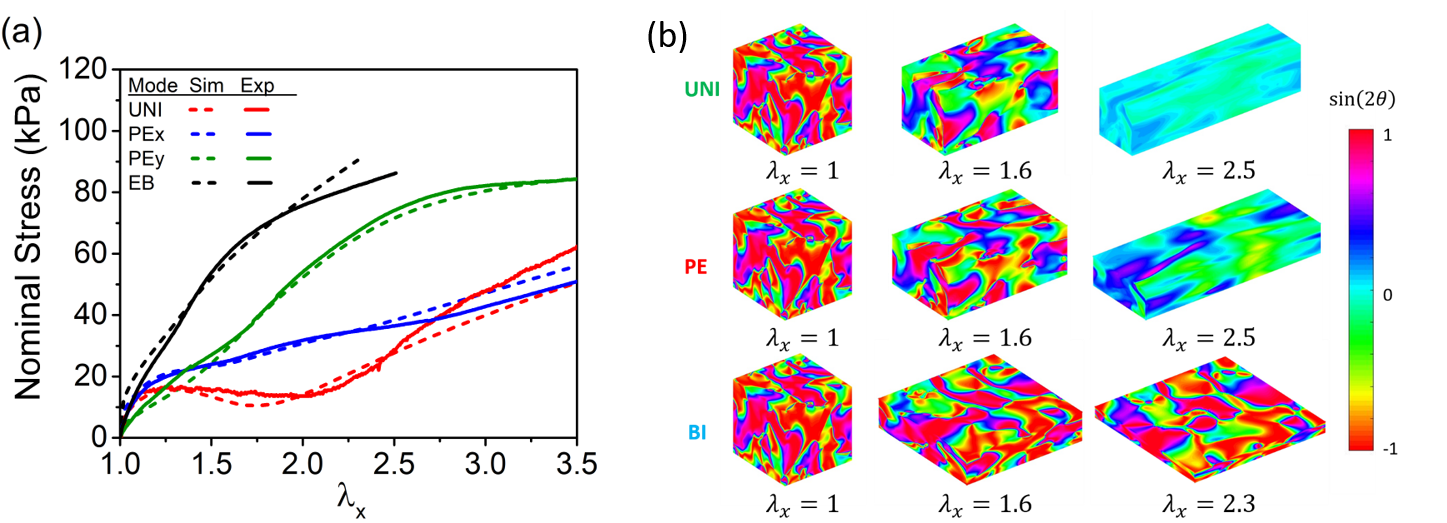}%
   		}
   		\vspace{-0.3in}
   	\end{center}
   	\caption{Unusual behavior of polydomain liquid crystal elastomers. (a) Nominal stress versus stretch under different loading conditions. (b) Microstructure evolution. ($\theta$ is the angle of $n$ with $x-$axis).}
   	\label{stress_stretch}
\end{figure*}

Recent experiments have suggested a peculiar behavior in LCEs.  When a sheet is stretched in planar extension (PE) where one in-plane direction is stretched while the other is held fixed ($\lambda_x>1, \lambda_y = 1$), the nominal stress in the stretched direction is smaller than that in the unstretched direction.  Figure \ref{stress_stretch}(a) shows the measured and computed stress-strain curves under uniaxial extension (U), PE and equibiaxial extension (EB) while Figure \ref{stress_stretch}(b) shows how the directors evolve.  Note that the simulations reproduce the complex observed behavior with a few parameters. More importantly, the simulations revealed that the microstructure evolves in such a way in PE that the in-plane shear was always zero, and this is the reason for the observed unusual behavior.  Other complex loading protocols also reveal good agreement.  Further, the statistics of microstructure evolution measured through X-ray scattering agrees well with the simulations.  In short, these simulations provide unique insight into microstructure evolution in LCEs.

The numerical performance and the scaling is demonstrated using the example of the elastic composite.  
First, the convergence with mesh is investigated. Taking the finest mesh as the reference, we calculate the relative error ($L_2$ norm) of deformation gradient, stress and displacement with different mesh size. They converge with an convergence rate of $1.83$, $1.84$ and $2.15$ respectively, close to the theoretically expected value of $2$.
Turning now to the scaling, we observe a steady decrease of wall time with increased threads of GPU. The slope is fitted as $-0.73$, suggesting very good scalability of the algorithm.  However, it is not perfect ($-1$) since FFT does not scale perfectly.  The scaling is also confirmed by weak scaling. The same configuration is studied using different mesh with proportional threads. Overall, the algorithm and GPU implementation show a good parallel efficiency as the system grows.

\begin{figure*}
   	\begin{center}
   		\resizebox{1.0\textwidth}{!}{%
   			\includegraphics[height=3cm]{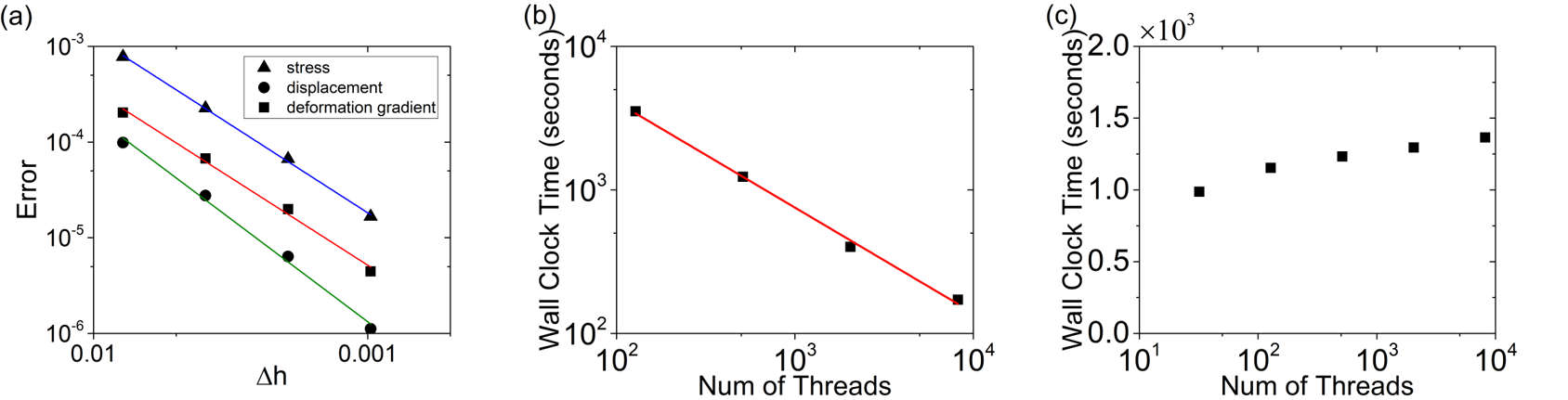}
   		}
   		\vspace{-0.5in}
   	\end{center}
   	\caption{Performance of accelerated computing.  (a) Relative error of physical quantities versus mesh size. (b) Strong scaling. (c) Weak scaling.}
   	\label{Scaling}
\end{figure*}

%%%%%%%%%%%%%%%%%%%%%%%%%%%%%%%%%%%%%%%%%%%%%
\section{Machine-learning material behavior} \label{sec:ml}

In this section, we describe a recent work where machine learning is used to link scales.   Machine-learning and especially deep neural networks have been extremely successful in image recognition \cite{lecun1995convolutional, he2016deep} and natural language processing tasks \cite{goldberg2017neural, collobert2008unified}.   There is also a growing literature on the use of these methods in materials science \cite{kd_armr_15}.  Machine learning has been combined with theoretical calculations, combinatorial synthesis, and high through-put characterization to rapidly identify materials with desired properties \cite{l_ncm_19,umehara2019analyzing,jetal_aplm_13}.
It has also been applied to parameter passing \cite{marchand2020machine, cole2020machine,wen2019hybrid} and 
to the inversion of experimental data \cite{m_mse_20,cd_mse_18}.
Image classification and natural language processing have been applied to approximate material constitutive behavior \cite{Mozaffar2019DeepPlasticity, jordan2020neural} 
and  homogenization of material behavior \cite{liu2019deep1, xiao2019machine}.

\begin{figure}
\centering
\includegraphics[width=4in]{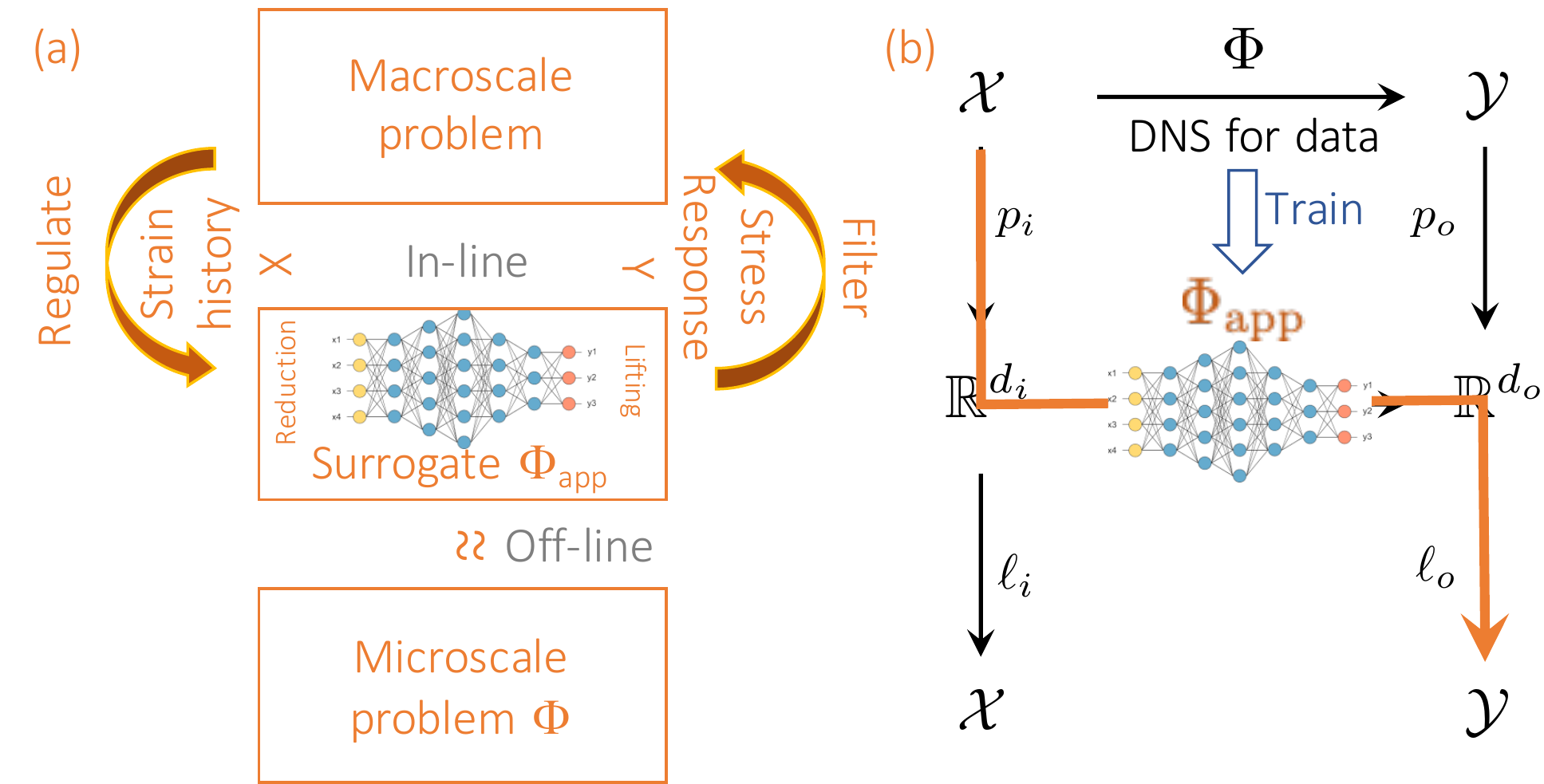}
\caption{Machine learning for scale transitions.  (a) Approach. (b) Surrogate construction.  \label{fig:approach}}
\end{figure}

Consider the two-scale setting shown in Figure \ref{fig:approach}a with the two scales labelled `microscopic' and `macroscopic'.  For definiteness, one may think of the macroscopic scale to be an application modeled using the finite element method and the microscopic scale to be a polycrystalline ensemble modeled using crystal plasticity.   The macroscopic model \emph{regulates} the microscopic model $\Phi$ by providing the deformation history of the macroscopic quadrature point, and this is the boundary condition in the microscopic model.    The resulting stress distribution in the microscopic model is then averaged or \emph{filtered} to return the required stress to the macroscopic model.  This framework can be justified using the homogenization method, and extended to any pair of scales in the multiscale modeling hierarchy.
However, the practical application of this approach is challenging.  We can solve the microscopic problem at each quadrature point at each instant of time, but this is computationally forbidding.  Or we can model the microscopic problem but this requires additional information thereby defeating the purpose of multiscale modeling.  So we seek to construct a surrogate $\Phi_{app}$ by learning the full solution operator of the microscale problem with no further empirical or expert input following \cite{letal_arxiv_21}.  
%We note that this is distinct from, and significantly more ambitious, accurate and fundamental compared to the other ongoing attempts that use ML as a tool for finding the empirically postulated constitutive response.

A critical challenge in the application of machine learning is that material models are described as partial differential equations (e.g. equilibrium and evolution equations in our demonstrative example) that map inputs from one function space (correspondingly, average strain history) to outputs on another function space (correspondingly, the stress response).   However, typical neural network architectures approximate finite dimensional inputs to finite dimensional outputs.  So a direct application makes the network dependent on the discretization which introduces artifacts and does not allow us to collect data from diverse sources.   One approach to overcome this is shown in Figure \ref{fig:approach}(b) and combines model reduction and neural networks for high-fidelity discretization-independent approximations of maps between function spaces \cite{Bhattacharya2020ModelPDEs}.  Briefly, we build an approximation of an identity map from a function space to itself by a composition of a model reduction map that maps from the function space to a finite dimensional space of latent variables and a lifting map that maps the finite dimensional space to the function space.  Examples of such maps are principal component analysis (PCA) and auto-encoders.  We implement this map to both the input and output spaces, and a neural network approximation between the finite-dimensional representations (latent variables).  The surrogate is a composition of the model reduction map in input space, the neural network approximation and lifting map to output space, and trained using data by the direct numerical simulation of the microscopic problem.
Importantly we are able to control this approximation in elliptic partial differential equations with theorems that bound the error in terms of the reduced dimensions, neural network features and training data; similar results are to be expected for parabolic problems..  Other approaches of machine learned approximations to the solution operator are explored in Li {\it et al.} \cite{zetal_neurips_20,zetal_iclr_21}.

\begin{figure}[t]
\centering
\includegraphics[width=5in]{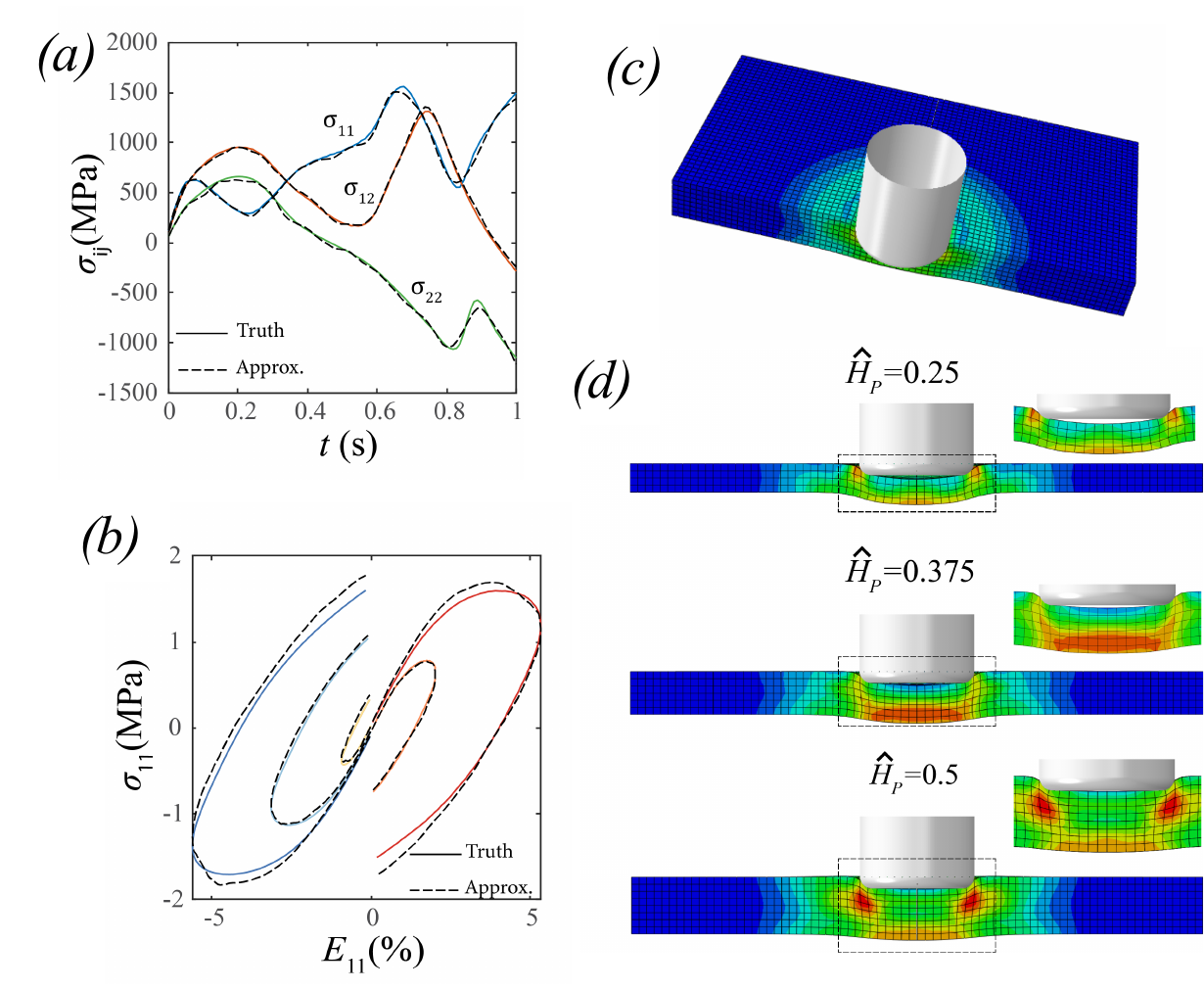}
\vspace{-0.3in}
\caption{Demonstration on impact of a polycrystalline material plate. (a,b) Predictive ability of machine-learned surrogate for deformation of polycrystalline Mg. (c) Ballistic plate impact and (d) design study of plate thickness using machine-learned surrogate.  \label{fig:impact}}
\end{figure}
We have demonstrated the approach to the problems of impact of polycrystalline magnesium in Figure \ref{fig:impact} \cite{letal_arxiv_21}.  High fidelity crystal plasticity unit cell calculations were used to create a machine learned surrogate that accurately predicts the material response against various strain histories (e.g. Figure \ref{fig:impact}(a,b)).
The surrogate is implemented as the material model (VUMAT) in the macroscopic solver (ABAQUS) to the study of a plate being impacted with a rigid, massive impactor Figure \ref{fig:impact}(c).  Once trained, the same surrogate can be used for a range of calculations including the design study on plate thickness (Figure \ref{fig:impact})(d) which shows a change of deformation mode from one dominated by bending to one dominated by punching) as well as the Taylor anvil test.  Importantly, a representative  calculation takes 2362 seconds, a few times larger than 262 seconds required for a similar calculation using an empirical constitutive relation (Johnson-Cook) and orders of magnitude smaller than 3.9 $\times 10^8$ seconds required for a concurrent calculation.  Furthermore, the calculation has all the physics of the concurrent calculation.  The one-time off-line cost of generating data (5.9 $\times 10^6$) and training (6.0 $\times 10^4$) are also smaller than a single concurrent calculation.   

%We have also used this approach to density functional theory to learn how the electronic structure changes with deformation\footnote{Y.S.\ Teh, S.\ Ghosh and K.\ Bhattacharya, Machine learnt electronic structure quantities as predictors for massive DFT calculations, In preparation}.

%%%%%%%%%%%%%%%%%%%%%%%%%%%%%%%%%%%%%%%%%%%%%
\section{Uncertainty quantification across scales} \label{sec:uq}

A key aspect of the materials-by-design approach is the recognition that properties of real materials are the result of complex multiscale phenomena. The complexity of the material behavior across scales is the main source of uncertainty in engineering applications. However, the direct estimation of integral material uncertainties requires repeated evaluations of integral material response aimed  at  determining  worst-case  scenarios  at  all  scales  resulting  in  the  largest  deviations  in  macroscopic behavior.  Such integral calculations are prohibitively expensive in terms of computational cost and are beyond the the scope of the present-day computers. 

We have developed a framework to quantify the propagation of uncertainties through length scales \cite{sun2020rigorous,liu2021hierarchical}. Crucially, no integral calculations are required at any stage of our analysis and the method only requires analysis of the unit mechanisms at each scale which is computationally feasible. Our approach supplies rigorous upper bounds of integral uncertainties for hierarchical multiscale material systems through a systematic computation of moduli  of continuity for each individual subsystem \cite{topcu2011rigorous}. The moduli of continuity provide the right measure for the interaction between the subsystems resulting in rigorous uncertainty bounds that are guaranteed to be conservative and applicable to safe designs. Unlike previous work, the framework is based on McDiarmid's inequality \cite{mcdiarmid1989method} and \emph{does not require a prior knowledge on the statistics of the material uncertainties} as in the cases for Bayesian based methods \cite{dashti2011uncertainty}. Furthermore, the bounds become sharper with an increasing number of input variables which is known as the concentration-of-measure phenomenon~\cite{ledoux2001concentration}. In addition, the computation of the bounds is non-intrusive and can be carried out using existing deterministic models of the subsystems and external scripts. 

\begin{figure}
\centering
\includegraphics[width=6in]{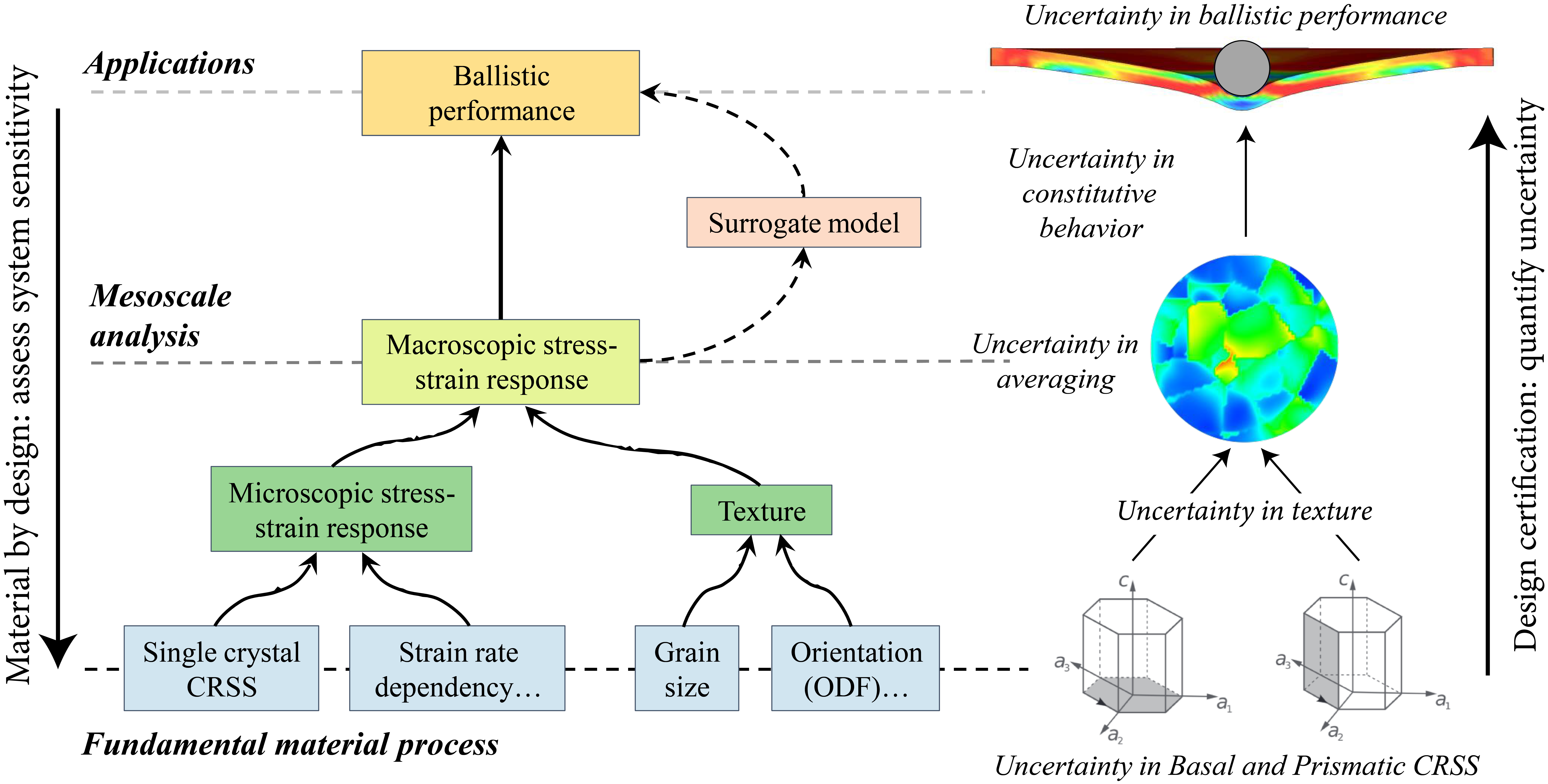}
\caption{Hierarchical multiscale uncertainty quantification of magnesium. \label{fig:HUQ}}
\end{figure}

\begin{figure}[t]
\centering
\includegraphics[width=5in]{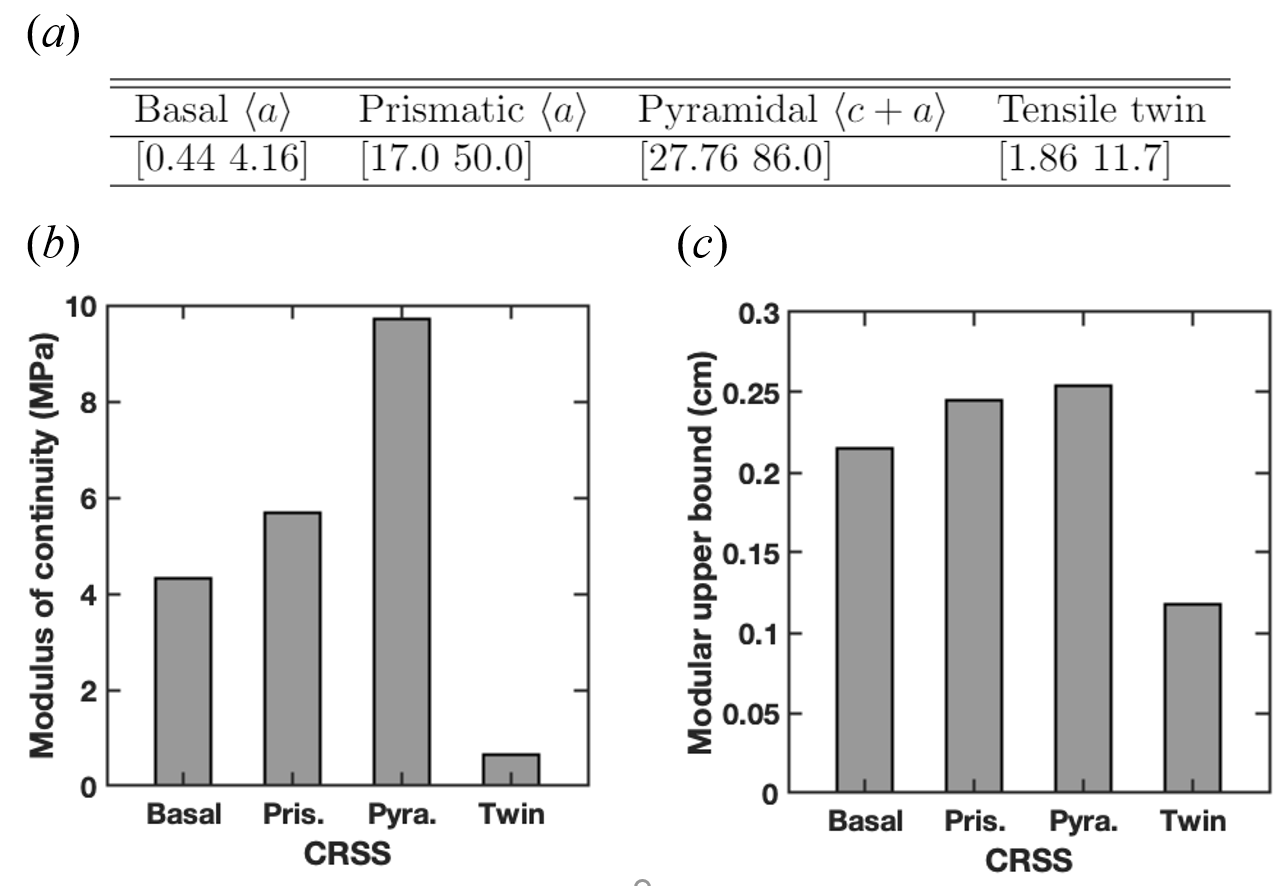}
\caption{Multiscale uncertainty of Magnesium plate. (a) Uncertainties in slip and twin critical resolved shear stress (CRSS). Unit: MPa. (b) Micro to meso uncertainty quantification showing the variation in Johnson Cook parameter from different CRSS. (c) Micro to macro uncertainty quantification showing the variation in backface deflection (modular upper bound) from different CRSS.  \label{fig:HUQdata}}
\end{figure}

The developed framework has been used to assess the ballistic impact of magnesium plate, as depicted in Fig. \ref{fig:HUQ}. Three scales of material behavior are considered here: (i) the microscale, where the behavior of the magnesium, including slip and twinning, is modeled
at the single crystal level; (ii) the mesoscale, where the polycrystalline response is computed using Taylor averaging and the single-crystal model; and (iii) the macroscopic impact problem, where the material behavior is approximated by the Johnson-Cook constitutive model and the ballistic performance of the magnesium plate is simulated using finite elements. We then study the uncertainty of the ballistic response due to uncertainties in the strength of individual slip and twin systems. In particular, we assume all the micro-scale parameters are uncertainty free except the slip and twin critical resolved shear stress (CRSS)~\cite{liu2021hierarchical}, see Fig.~\ref{fig:HUQdata}(a). The micro-scale uncertainty results in the uncertainties of the meso-scale Johnson Cook parameters and eventually the macroscale back-face deflection (quantity of interest).

The resultant micro to meso uncertainty propogation and micro to macro uncertainty propagation are demonstrated in Fig. \ref{fig:HUQdata} (b) and (c) respectively. An important property of the moduli of continuity is that, since they are dimensionally homogeneous, they can be compared and rank-ordered, which in turn provides a quantitative metric of the relative contributions of the input parameters to the overall uncertainty. The rank-ordering of the CRSSs to the overall uncertainty in ballistic performance is found to be pyramidal$>$prismatic$>$basal$>$twin, with the pyramidal and prismatic CRSSs contributing the most, the twin CRSS the least and the basal CRSS in between. The
calculations show that the integral uncertainties determined by the hierarchical multiscale UQ approach are
sufficiently tight for use in engineering applications. The analysis also sheds light on the relative contributions of the different unit mechanisms (i.e., slip and twin systems) to the integral uncertainty and the dominant propagation paths for uncertainty across the model hierarchy.

%%%%%%%%%%%%%%%%%%%%%%%%%%%%%%%%%%%%%%%%%%%%%
\section{Data-driven simulations} \label{sec:data}

The computational sciences, as applied to physics and engineering problems, have always been about using data inputs to predict outcomes. Computational science differs from generic data science (cf., e.~g., \cite{WD001, WD002, WD003, WD004}) in that the problems of interest are constrained by  physical laws. Such physical laws find mathematical expression in the form of field equations such as conservation of linear momentum in mechanics, conservation of mass in transport problems, Maxwell's equations in electromagnetism, and others. Many of the methodologies that have been developed since the dawn of modern numerical analysis in the 1950's have been preoccupied with discretizing those exactly-known field equations. Finite differences, finite elements, finite volumes, molecular dynamics and mesh-free methods are all examples of different ways of approximating field equations.

By contrast, material laws have traditionally been uncertain and imperfectly known owing to the paucity of observational data, experimental error or intrinsic stochasticity of the material behavior. A common response to this imperfect knowledge has been modeling. Models are designed to act as succinct summaries of complex material data for which summarization is a difficult task. They are also used to characterize material behavior across regimes that are data sparse. To that end, material modeling relies on heuristics and intuition, which inevitably results in loss of information, biasing, modeling error and epistemic uncertainty. Adding to these difficulties, material modeling is open-ended, i.~e., there is no theory that dictates how sequences of models of increasing accuracy and fidelity can be generated that are sure to converge to the actual--and unknown--material law. 

For instance, a common approach to material modeling is to assume a relation of the form
\begin{equation}\label{zkmkcf}
    \sigma = g(\epsilon) + \eta ,
\end{equation}
where $\epsilon$ and $\sigma$ are local work-conjugate variables characteristic of the material, $g$ is a deterministic material law and $\eta$ represents observational noise. In practice, the essential difficulty is that neither $g$ nor the distribution of $\eta$ are known. General principles do not suffice to determine $g$ uniquely with material specificity and considerable latitude is left to the modeler as regards material identification. A common approach is to determine $g$ by recursion and fitting to empirical data, e.~g., by means of machine learning. Here again, considerable latitude is left to the modeler as regards the type of functions and criteria to be used for purposes of recursion and representation. For stochastic systems, the situation is greatly compounded by the fact that the prior probability distribution from which the noise $\eta$ is drawn is generally not known and needs to be modeled as well.

Against this backdrop, the staggering developments in experimental science and microscopy, which produce large sets of fully-resolved 3D material data, constitute a veritable game changer. By virtue of those advances, scientific computing has transitioned from being data-poor to being data-rich. This sea change raises the question of whether a more direct connection between material data and prediction can be effected, specifically one where the intervening modeling step is eliminated altogether. A notional comparison between classical and model-free inference is
\begin{equation*}
\begin{array}{cccccc}
    \text{Classical inference:}
    & \text{Data} & \to & \text{Model} & \to & \text{Prediction}
    \\
    \text{Model-Free Data-Driven inference:}
    & \text{Data} &  & \longrightarrow &  & \text{Prediction}
\end{array}
\end{equation*}
Here, modeling is understood as any operation that modifies the data set or replaces it by another object, be it through constitutive modeling, fitting, model reduction, regression, machine learning, or any other operation. By model-free we understand methods of inference that use the data, all the data and nothing but the data for purposes of prediction. Material behavior is explicitly defined by the source data associations and modeling empiricism, error and uncertainty are eliminated entirely.

There is considerable ongoing work aimed at developing this emerging Model-Free Data-Driven paradigm. The initial proposal \cite{WD005} developed a distance-minimizing approach that converges for sequences of uniformly convergent data sets. Subsequent work \cite{WD005a} made use of a max-ent information theory in order to render the approach robust with respect to noise and outliers in the data set. The distance minimization approach has also been extended to infinite-dimensional boundary-value problems, including linear and finite elasticity \cite{conti2018a, conti2020a, Platzer:2021}. This extension differs fundamentally from the finite-dimensional setting in that relaxation, or the emergence of weakly convergence fine spatial oscillations that exploit the structure of the data set, plays an essential role. Remarkably, relaxation sets forth a notion of convergence with respect to data, or $\Delta$-convergence that is fundamentally different from the classical relaxation of energy functionals. Extensions of Model-Free Data-Driven analysis to dynamics and inelastic materials require consideration of evolving data sets conditioned by the prior history of the material \cite{Kirchdoerfer:2018, eggersmann2019a, carrara2020a}.

The Model-Free Data-Driven paradigm also sets forth a novel alternative to calculus of variations, concurrent and parameter-passing schemes in multiscale modeling. 
Specifically, one can use offline microscale calculations in order to generate material data sets for Model-Free Data-Driven computing at the macroscale.  This is a fully automated approach that requires no modeling or analysis at the macroscale and requires no fitting of the micromechanical data. The approach is lossless in that it uses the generated microschanical data, all the data and nothing but the data.

\begin{figure}
\begin{center}
    \includegraphics[width=0.99\linewidth]{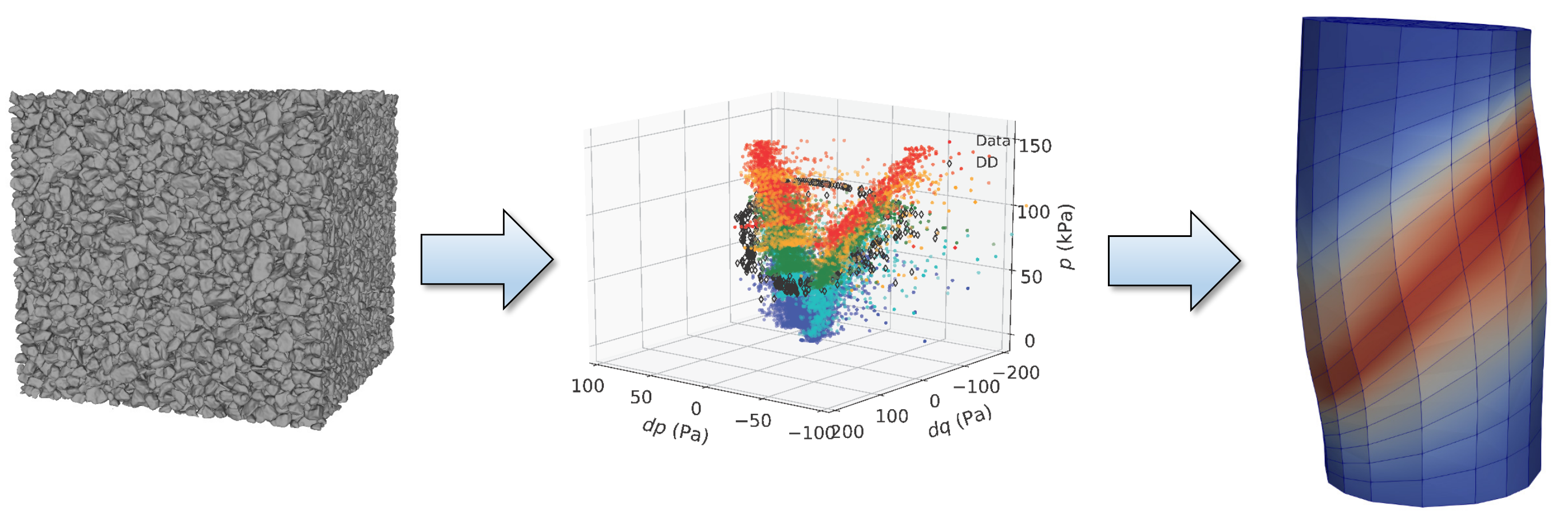}
    \caption{Model-Free Data-Driven multiscale analysis of plane-strain uniaxial compression test of \cite{karapiperis2021a}. Left: Representative volume element for micromechanical calculations. Middle: Sample data computed by deforming RVE along selected strain paths using the Level-Set Discrete Element Method (LS-DEM) \cite{kawamoto2016a}. Right: Model-Free Data-Driven finite-element calculation of a triaxial compression experiment on a specimen of angular (Hostun) sand \cite{Ando2012} built on the RVE data.} \label{wJ0N9x}
\end{center}
\end{figure}

Karapiperis {\sl et al.} \cite{karapiperis2021a} have demonstrated the approach by means of an application to granular media. Specifically, they rely on the Level-Set Discrete Element Method (LS-DEM) \cite{kawamoto2016a} to generate material data sets for angular (Hostun) sand, Fig.~\ref{wJ0N9x}. The virtual specimen is deformed along selected loading and unloading paths to produce material data sets including stress, strain, internal energy and dissipation, in accordance with an energy-based parametrization of the data. A representative data set is shown in Fig.~\ref{wJ0N9x}. The data thus obtained can be used, {\sl in toto}, as a basis for Model-Free Data-Driven calculations at the macroscale. Fig.~\ref{wJ0N9x} depicts one such simulation of an \textit{in-situ} triaxial compression experiment on a specimen of angular (Hostun) sand~\cite{Ando2012}. The specimen is first compressed isotropically to 100 kPa, and then compressed triaxially by keeping the cell pressure constant while prescribing a vertical displacement to the platen under quasistatic conditions. Failure is computed to occur through the formation of a persistent shear band, which agrees well with experimental observations \cite{Ando2012}. The axial strains, principal stress ratio and volumetric strains predicted by the Model-Free Data-Driven simulation are also in good agreement with experimental measurements.

It bears emphasis that, in this approach, the micromechanical data is lifted into the macromechanical calculations losslessly and without modification, i.~e., without modeling of any type. This direct link between data and prediction effectively cuts through the Gordian knot of multiscale upscaling and the representation of the effective macroscopic behavior. 

Data-Driven methods are likely to gain importance at a time when data from high-fidelity simulations and high-resolution experiments are becoming increasingly abundant. The Model-Free Data-Driven paradigm, in particular, possesses ancillary attributes that add to its appeal. Thus, it standardizes solvers by separating the treatment of the field equations from the characterization of material behavior. Specifically, Model-Free Data-Driven computing reduces boundary value problems to the solution of two standard {\sl linear} problems, regardless of material behavior. The interaction between the solver and the material data repository reduces to data searches and data transfer that can also be standardized and scripted non-intrusively. In particular, the data repositories can be centralized, developed and maintained remotely from the locally run solver software. This data and work-flow structure allow material data sets of disparate provenances to be pooled together and has the potential for changing the way in which material data is developed, stored, exchanged and disseminated in science and in industry.

%%%%%%%%%%%%%%%%%%%%%%%%%%%%%%%%%%%%%%%%%%%%%
\section{Application-driven materials optimization} \label{sec:app}
The performance of modern devices typically depends on the properties of their component materials that operate across a broad range of disparate time- and length-scales~\cite{sun2017acceleration, sun2019atomistic, zheng2016multiscale}. Due to recent advancements in nanotechnology, materials characterization and synthesis, additive manufacturing, and high-performance computing, it is possible to develop innovative advanced materials with sophisticated structures and multiple properties from the atomistic to the macroscopic application scale \cite{sun2018long, pikul2019high, meza2014strong}. For instance, mechanical properties can be tailored by adjusting microstructures, material constitutions, their spatial distribution and mass fractions using material synthesis and processing \cite{yogeshvaran2020out, bhattacharya2003microstructure,liu2019deep}. On the other hand, structural properties, e.~g., size and shape of components, can be controlled by material manufacturing such as 3D printing at different length-scales \cite{wong2012review, schaedler2011ultralight, schumacher2015microstructures}. 

Traditionally, a device can be designed in a {\it sequential} manner. First, each individual component material is optimized over its desired design properties separately in very simple tests, or even by \emph{serendipitous} discoveries. Then, the device, as a whole system, is designed with respect to the rest of the properties, e.~g., the distribution and fraction of component materials in the device, whereas the properties of each component are fixed at the ``optimal'' results determined from the previous step. However, there exist two critical challenges in this sequential optimization process. First, the attainment of all the best properties is a vital requirement for most materials; unfortunately these properties are generally mutually exclusive, e.~g., the conflict between strength and toughness. Therefore, how to account for the correlations between design properties of component materials poses a significant challenge to the development of new devices in the applications of interest. Second, the sequential process neglects the connection between component materials, and the connection between the entire device and its application environments. As a result, the performance of the device might be underestimated over the accessible ranges of its design properties. 

\begin{figure}
\centering
\includegraphics[width=6.5in]{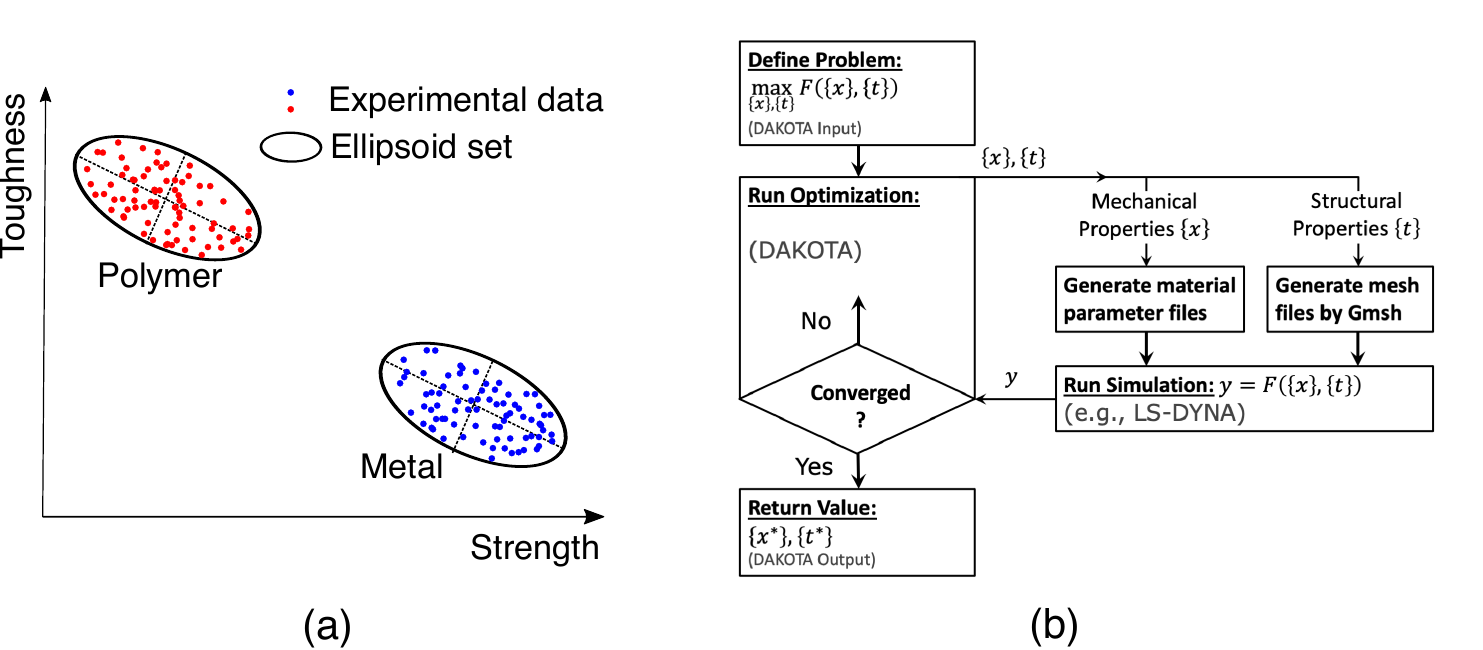}
\caption{(a) Schematic illustration of conflict between mechanical properties. (b) Flowchart of joint design strategy over mechanical and structural properties.}
\label{fig:flowchart}
\end{figure}

In order to address the aforementioned problems, we have proposed a {\it joint} design strategy~\cite{sun2021goal}. The basic idea behind this strategy is that, the device with its components is regarded as a whole system and the design process is directly related to the application where the device operates. To this end, the design variables include both the properties of each individual component material and the properties connecting the components with the device. The design objective is to maximize a crucial metric that characterizes the targeted performance of the device in the application. At the same time, the correlations between material properties are taken into account by employing ellipsoid convex sets as inequality constraints in the optimization~\cite{jiang2011correlation, jiang2013structural}, as illustrated in Fig.~\ref{fig:flowchart}(a). As a result, the design of the device is formulated as a constrained co-optimization problem that is solved over all the design parameters \emph{simultaneously}. We specifically consider the mechanical and structural design variables of the device and have developed a non-intrusive, high-performance computational framework, Fig.~\ref{fig:flowchart}(b), based on DAKOTA Version $6.12$ software package~\cite{adams2020dakota} of the Sandia National Laboratories. We have also implemented Gmsh Version $4.5.4$ software package~\cite{geuzaine2009gmsh} in the framework, since the optimization requires evaluation of different structural parameters and therefore may need to generate meshes for the device on-the-fly.

\begin{figure}
\centering
\includegraphics[width=6.5in]{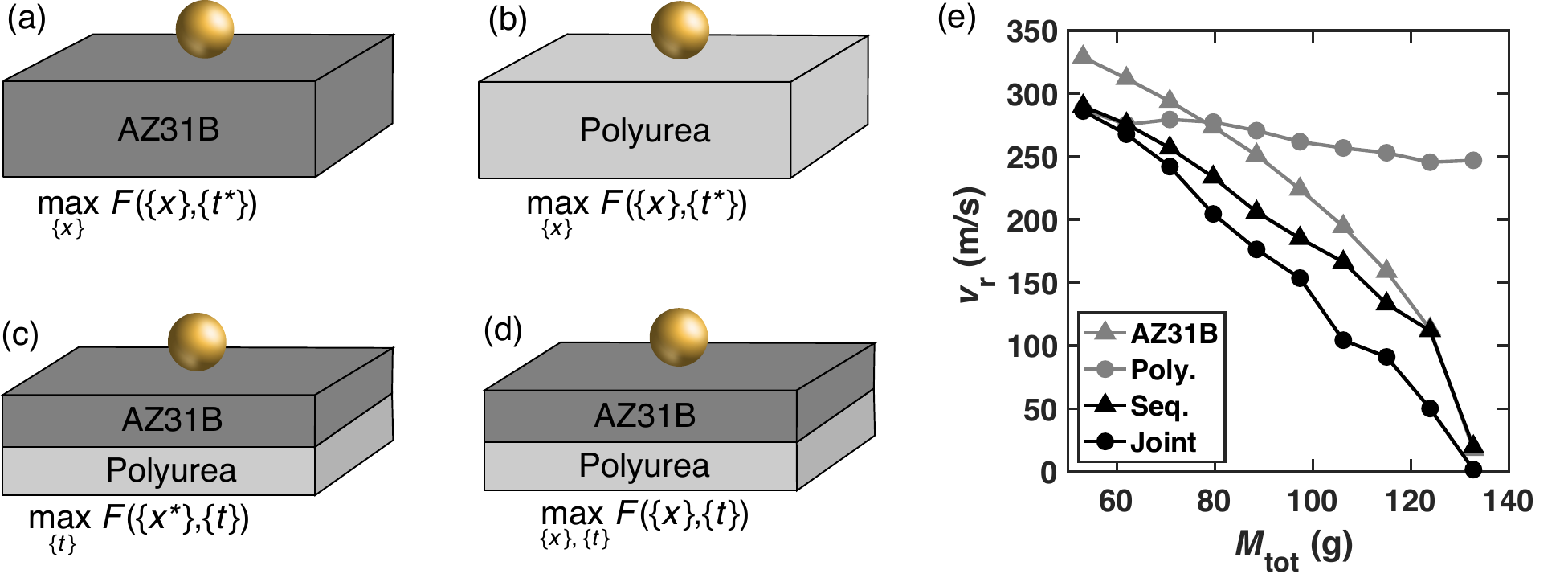}
\caption{Strategies and results of designing a multi-layered plate subject to high-speed impact. (a) Design over mechanical properties of AZ31B. (b) Design over mechanical properties of polyurea. (c) Design over structural properties of layers. Steps (a), (b) and (c) form the strategy of sequential design. (d) Joint design over mechanical and structural properties. (e) Comparison of residual velocity at different values of plate mass.}
\label{fig:case}
\end{figure}

We assess the sequential and joint design strategies in an application where the ballistic performance of a double-layered plate is optimized using alternating AZ31B magnesium alloy and polyurea, as depicted in Fig.~\ref{fig:case}. Schematic illustrations of the two design strategies in this application are shown in Figs.~\ref{fig:case}(a)-(d). Specifically, we consider a scenario in which normal impact and full perforation take place during the impact process. The design objective is assumed to be a minimum residual velocity $v_\text{r}$ of the projectile after penetrating the plate. The design variables include the mechanical properties of AZ31B and polyurea that govern the strength and toughness of materials, and the structural parameters of component layers, i.~e., their thicknesses. The constraints include a fixed total mass of the plate $M_\text{tot}$ and ellipsoid convex sets on the mechanical properties of AZ31B and polyurea. Fig.~\ref{fig:case}(e) compares the residual velocity at the optimal mechanical and structural parameters using different values of $M_\text{tot}$. Notably, using residual velocity as a reference, the proposed joint design strategy can greatly improve the performance at the intermediate values of the plate mass compared to the traditional sequential design approach. Importantly, our result not only agrees well with the classical knowledge in ballistic protection systems that a combination of strong and stiff front layer with a soft but tough back layer provides the optimal ballistic performance \cite{o2014mechanisms, liu2019failure,liu2019high}, but also provides a quantitative optimal solution of the design. Therefore, the proposed method has provided new insights for designing novel materials with significant desired application performance.

%%%%%%%%%%%%%%%%%%%%%%%%%%%%%%%%%%%%%%%%%%%%%
\section{Summary}

In this paper, we have reviewed a number of recent developments in methodology that advance the goal of designing materials targeted by specific applications.  The macroscopic behavior of materials is the end result of  a number of mechanisms that operate across a broad range of disparate scales.  Multiscale modeling seeks to address this complexity using a `divide and conquer' approach where range of material behavior is first divided into an ordered hierarchy of scales.  Section \ref{sec:acc} describes how emerging computational platforms that use accelerators like GPUs can be effectively used to study problems at individual scales.  The scales are then linked using an approach where the higher scale model modulates the lower scale behavior, but also averages its outcome.  Section \ref{sec:ml} shows how machine learning can be effectively used for this scale transition.  The complexity of multiscale modeling also introduces uncertainty at various scales.  Section \ref{sec:uq} shows how we can exploit the hierarchical nature of multiscale modeling to quantify integral uncertainties by studying individual uncertainties.  Section \ref{sec:data} describes an entirely new approach of model-free simulations where the data -- experimental or computational from lower scale models -- can be exploited directlly.  Finally, we explore the joint optimization of material and structure in Section \ref{sec:app}.

%%%%%%%%%%%%%%%%%%%%%%%%%%%%%%%%%%%%%%%%%%%%%
\section*{Acknowledgement}

The work described in Sections 3, 4 and 6 was sponsored by the Army Research Laboratory and was accomplished under Cooperative Agreement Number W911NF-12-2-0022. The views and conclusions contained in this document are those of the authors and should not be interpreted as representing the official policies, either expressed or implied, of the Army Research Laboratory or the U.S. Government. The U.S. Government is authorized to reproduce and distribute reprints for Government purposes notwithstanding any copyright notation herein
The work described in Section 2 was sponsored by the Air Force Office of Scientific Research under the MURI Award FA9550-16-1-0566.  The work described in Section 5 was sponsored by the  Air Force Office of Scientific Research through the Center of Excellence on High-Rate Deformation Physics of Heterogeneous Materials,  Award FA9550-12-1-0091 and the Deutsche Forschungsgemeinschaft through the Sonderforschungsbereich 1060 `The mathematics of emergent effects'.

%Bibliography
%\newpage
\bibliographystyle{abbrv}
\bibliography{materials_systems}

\end{document}